# ANALYTICAL MODEL FOR MOBILE USER CONNECTIVITY IN COEXISTING FEMTOCELL/MACROCELL NETWORKS

Saied M. Abd El-atty[1] and Z. M. Gharsseldien[2]

[1]Department of Computer Science and Information, Arts and Science College, Salman Bin Abdulaziz University, 54-11991,Wadi Adwassir, Kingdom of Saudi Arabia
`s.soliman@sau.edu.sa`
[2]Department of Mathematics, Arts and Science College, Salman Bin Abdulaziz University, 54-11991,Wadi Adwassir, Kingdom of Saudi Arabia
`z.gharsseldien@sau.edu.sa`

## ABSTRACT

*In this paper we investigate the performance of mobile user connectivity in femtocell/macrocell networks. The femto user equipment (FUE) can connect to femto access point (FAP) with low communication range rather than higher communication range to macro base station (MBS). Furthermore, in such emerging networks, the spatial reuse of resources is permissible and the transmission range can be decreased, then the probability of connectivity is high. Thereby in this study, we propose a tractable analytical model for the connectivity probability based on communication range and the mobility of mobile users in femtocell/macrocell networks. Further, we study the interplays between outage probability and spectral efficiency in such networks. Numerical results demonstrate the effectiveness of computing the connectivity probability in femtocell/macrocell networks.*

## KEYWORDS

*Femtocell, Macrocell, Connectivity, Mobility, Communication range.*

## 1. INTRODUCTION

Integration of femtocell technology with the existing macrocell mobile networks is a promising solution not only to improve indoor coverage but also to increase capacity of cellular mobile networks. Therefore, the deployment of femtocells technology will be useful for both users and mobile network operators, since femtocell networks introduce better quality wireless services and data transmission. Furthermore, the users make use of femtocell networks to receive strong signals, and high capacity, as well as low transmission range and power wasting [1]. As a result, the mobile network operators will have the solutions for radio resources limitations, reduction macrocell traffic load and infrastructure cost by saving lots of money by offloading most of the capital expenditure (CAPEX) and operational expenditure (OPEX) onto users [2].

In femtocell networks, the smaller size of femtocell not only provides high spectrum efficiency by using spatial reuse of resources but also decreases the transmission range and then provides high probability of connectivity.

---

The permanent address of the authors is
[1]The Dept. of Electronics and Electrical Communications,
 Faculty of Electronic Engineering, Menoufia University, 32952, Menouf, Egypt.
[2]The Dept. of Math., Fac. Sci., Al-Azhar Uni., Nasr City,11884, Cairo, Egypt

             61

Further, one of the most important reasons for deployment femtocells technology is to serve the remote areas with no coverage or poor signal as well as to reduce the traffic volume on the macrocell. Hence in this study, we introduce a mathematical model for the probability of connectivity as a function of communication range and the mobility of mobile user in femtocell/macrocell network. As well as, in terms of connectivity probability, we study the interplays between the outage probability and the spectral efficiency based-signal to interference ratio (SIR) in such networks.

The rest of the paper is organized as follows. The related work which discussed the technical challenges associated with femtocell deployment is presented in Section 2. Femtocell access methods comparisons are investigated in Section 3. Then, the main beneficial of femtocell technology deployment and the mathematical model for mobile user connectivity are introduced at section 4. Femtocell/macrocell network modeling and the interaction between outage probability and spectral efficiency are presented in Section 5. In Section 6, the numerical results are discussed. Finally, the paper is concluded at Section 7.

## 2. RELATED WORK

Most of studies and researches in Femto-Macro cellular networks are focused on proposing the schemes to overcome the technical challenges in such networks such as, interference management, handover control, spectrum allocation, access methods and etc. In [3], the authors proposed an efficient hybrid frequency assignment technique based on interference limited coverage area (ILCA). They studied two different scenarios of path loss for calculation ILCA. On the same direction the authors in [4] introduced a decentralized resource allocation scheme for shared spectrum in macro/femto OFDMA networks in order to avoid inter-cell interference. In addition, the authors in [5] exploited the femto size feature to propose a resource reuse scheme based on split reuse and graph theory. In sequel, the authors in [6] have proposed a radio resource management in self organizing femtocell and macrocell networks to satisfy the demanding of selfish users. Further, according to the proposed scheme, they have introduced incentives for femtocell users to share their FAPs with public users.

On the other hand, the authors in [7] studied the outage probability in macrocell integrated with femtocell CDMA networks. They concluded that femtocell exclusion region and a tier selection based handoff policy offers modest improvements in area spectral efficiency (ASE). The handover procedure in femtocell integrated with 3GPP LTE network is analyzed for three different scenarios: hand-in, hand-out and inter-FAP in [8]. Furthermore, the authors in [9] have proposed a novel handover decision algorithm according to the location of user in the femtocell coverage and at the same time taking into account the received signal strength (RSS) from femtocell and macrocell. In [10] the authors have investigated the existing access methods for femtocells with their benefits and drawbacks. They have also provided a description for business model and technical impact of access methods in femto/macro networks. Subsequently, a framework for femtocell access method of both licensed and unlicensed band with the coexisting with WiFi is introduced in [11]. In addition, the conflict between open and close access methods is compared in [12]. They focused on the downlink of femtocell networks and they evaluated the average throughput as a function of SINR of home and cellular users for open and closed access methods.

Unlike the above literature that mainly focus on introducing the solutions for challenges in femtocell/macrocell networks. To the best of our knowledge, our work is able to provide a convenient solution for computing the probability of connectivity in such networks. More specifically, we propose an analytical model in femtocell/macrocell networks in order to compute the probability of connectivity in terms of communication range, and mobility of mobile users. As well as we study the interplays between the probability of outage and spectral efficiency in such networks. Further, we consider a femtocellular network based-open access method (OAM) in order to enhance the macrocell users (MUEs) link reliability by selecting the closest FAPs.

## 3. FEMTOCELL ACCESS METHODS

### 3.1. Closed Access Method (CAM)

In this scenario, the FAP serves only the authorized users, therefore the macro users are not allowed to access FAP. As portrait in Fig.1, although the macro user MUE1 or MUE3 is close to the radio coverage of femtocell, they cannot access FAPs [1]. On the other hand, the femto users (FUEs) prefer CAM for private access in order to protect privacy [12]. In sequel, the femto users don't like to share the limited capacity of FAP with others if no cost revenue. However, deploying CAM causes severe cross-tier interference from macro users in the reverse link or to nearby macro users in forward link. Therefore, interference mitigation and spectrum management are technical challenges in CAM [2].

### 3.2. Open Access Method (OAM)

In order to reduce the cross-tier interferences the OAM is considered. In OAM any passing macro user can access a FAP if the macro user is within the radio coverage of femtocell. Alternatively the macro user MUE1 or MUE3 causes or experiences strong interference, they can access FAP as shown in Fig.2. Therefore, OAM is more efficient in improving system capacity because OAM is able to serve the macro users that causing interferences. However, OAM introduces an enormous number of handover requests and consequently high communication overhead on both the radio access networks (RAN) and IP core networks (CN) [1]. In our system model, we considered the OAM method since we would gain not only improving system capacity but also reducing interferences. On the other hand, the handover problems can be solved by handover control mechanism.

### 3.3. Hybrid Access Method (HAM)

As we have seen previously all access methods suffer from pros and cons. Hybrid access method (HAM) is considered an adaptive method between OAM and CAM [2]. In HAM a portion of FAP resources are reserved for exclusive use of the CAM and the remaining resources are assigned in an open manners; thereby the outage number of users in macrocell can be reduced. This procedure should be controlled to avoid high blocking probability in femtocell and to reduce the annoying of registered users [12]. Nowadays, Femto forum, Broadband forum and 3GPP specifications try to finish the most efficient access method in femtocell networks but it is still under scrutiny.

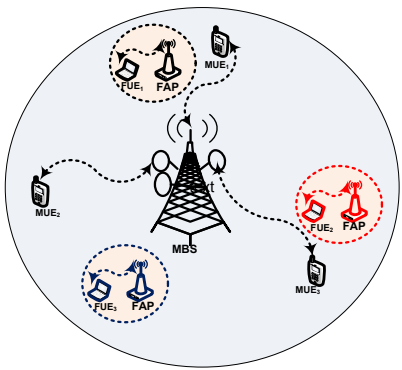

Figure 1. Closed access method (CAM) scenario.

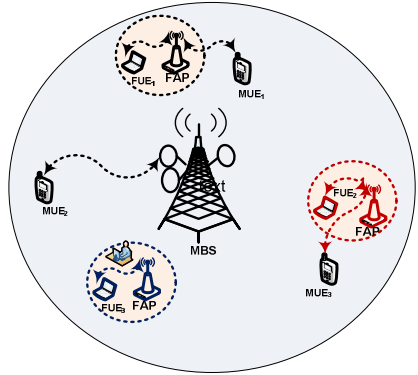

Figure 2. Closed access method (CAM) scenario.

## 4. BENEFITS OF DEPLOYMENT FEMTOCELLS TECHNOLOGY

Femtocells are self-organized networks (SONs) that are integrating itself into the mobile network without user intervention and then reducing deployment cost [20]. One of the key features of deployment the femtocells technology integrated with the current mobile cellular networks is that FUE or MUE requires no new equipment and hence femtocells technology does not require dual-mode handset [21]. Furthermore, the femtocell networks are capable of serving the remote areas with no coverage or poor signal as well as reducing the traffic volume on the macrocell. Therefore in terms of the communication range and the mobility factor of mobile users, in the following subsection, we introduce a tractable mathematical model for connectivity probability of mobile user.

### 4.1. Analytical Model for mobile user connectivity

We consider a scenario of a femtocells network; the mobile users are randomly and uniformly distributed in its coverage area. In addition, we assume a femtocells network is deployed within the communication range of a macrocell base station (MBS) and is allowed open access method (OAM) in order to enhance the mobile users (MUE) link reliability by selecting the closest FAPs.

We assume that the coverage area of femtocell is a circle with unity radius and a particular MUE with a communication range $r$ ($r$ <1) can access FAP by employing OAM. According to the mobility of mobile users, we have three cases to study
 • MUE is completely inside the femtocell range,
 • MUE is completely outside the femtocell range,
 • MUE communication range intersects with the femtocell range.

Let the distance between the center of FAP and the MUE is called $d$, which plays an important role in this analysis. There are crucial values for $d$ that may make the MUE is completely outside of the femtocell ($d=1+r$ or more) or the MUE is fully inside ($d=1-r$ or less) as shown in Fig.3. Accordingly, we can utilize a new parameter $\beta$ which expresses the area corresponding to the case of the MUE; it can be expressed as follows

$$d = 1 + \beta r \quad (1)$$

This parameter $\beta$ called the mobility factor of mobile user, and plays an important role to avoid the MUE from disconnectivity.

The disconnectivity defined as the probability of at least one MUE being out of coverage region of all other femtocells in a given macrocell. The probability of MUE being in disconnectivity may be determined by calculating the area of intersection between a circle of unity radius and circle of radius *r* as follows:

$$A(r,\beta) = \begin{cases} 0, & \beta \geq 1 \\ h(r,\beta), & -1 < \beta < 1 \\ \pi r^2 & \beta \leq -1 \end{cases} \quad (2)$$

In this scenario, the MUE is completely outside femtocell if $\beta \geq 1$, completely inside femtocell if $\beta \leq -1$, and partially intersected with femtocell if $1 < \beta < -1$ as illustrated in Fig.3. In order to find the area of the intersection between MUE communication range's and femtocell coverage's range as a function of *r* and $\beta$, the equations of the circles in Cartesian (*x*, *y*) plane are given by

$$x^2 + y^2 = r^2,$$
$$(x-d)^2 + y^2 = 1 \quad (3)$$

These circles intersect at

$$x_0 = \frac{r^2 + d^2 - 1}{2d} \quad (4)$$

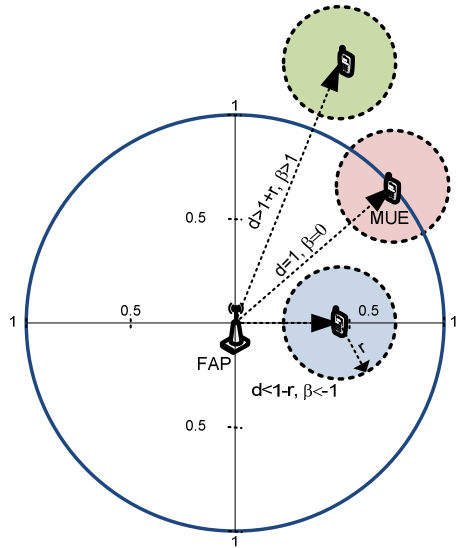

Figure 3. The mobility scenarios of mobile user in femtocell network.

By using the integration technique, we can obtain the function $h(r, \beta)$ as follows:

$$h(r,\beta) = \frac{\pi(1+r^2)}{2} - x_0\sqrt{r^2 - x_0^2}$$
$$- (d - x_0)\sqrt{1 - (d - x_0)^2}$$
$$- \sin^{-1}(d - x_0) - r^2 \sin^{-1}\left(\frac{x_0}{r}\right) \quad (5)$$

All femtocells are mutually exclusive through this analysis, and each MUE only intersects with one femtocell. Hence, the intersected area with $m^{th}$ femtocell may be written as follows:

$$A_j(r,\beta) = \begin{cases} A(r,\beta), & j = 1 \\ 0 & j \neq 1 \end{cases} \quad (6)$$

Assuming that there are $N_f$ number of femtocells overlaid in the macrocell mobile network. Hence, the probability of MUE is not able to connect to any femtocell among $N_f$ being ($P_j$) is given by the probability that all $N_f$ -1 femtocells lie in uncovered region. Then

$$P_j = \left(\frac{S - A_j(r,\beta)}{S}\right)^{N_f - 1} = \left(1 - \frac{A_j(r,\beta)}{S}\right)^{N_f - 1} \quad (7)$$

where $S = \pi$, $j=1, 2,…, N_f$ is the area of the circle with unity radius for all femtocells. Thereby, the probability of at least one MUE being disconnected ($P_d^1$) is given by $P_d^1 = \bigcup_{x=1}^{N_f} P_x$. Upper bound on $P_d^1$ can be obtained by using the union bound definition as

$$P_d^1 = \sum_{x=1}^{N_f} \left(1 - \frac{A_j(r,\beta)}{S}\right)^{N_f - 1} \quad (8)$$

By using (5)

$$P_d^1 \leq \left(1 - \frac{A(r,\beta)}{\pi}\right)^{N_f - 1} \quad (9)$$

Thereby, we can obtain the probability of connectivity ($P_C$) as the complement of disconnectivity probability of at least one MUE being isolated ($P_d^1$), hence from (9), we obtain

$$P_C \leq \left[1 - \left(1 - \frac{A(r,\beta)}{\pi}\right)^{N_f - 1}\right] \quad (10)$$

## 5. FEMTOCELL/MACROCELL NETWORK MODELING

We consider a network scenario of open access method (OAM) in the hierarchical macrocell with highly dense femtocells as depicted in Fig.4, where $N_f$ femtocells are uniformly distributed in the macrocell. The density of FAPs and the density of users are represented by $D_f$ (FAPs/m$^2$) and $D_u$ (users/m$^2$) respectively. We assume $D_u$ is a random stochastic Poisson process follows spatial distribution [13]. A simplified wireless channel model is considered in our network model (i.e., fading, shadowing, noise … etc. is neglected), since the channel model is a distance-dependent according to the path loss with exponent $\alpha$ >2. Subsequently, considering all users transmit with a fixed power $P_t$ and all FAPs have the same power level of sensitivity, $P_{min}$. Thereby, we can express the communication range $r$ in the reverse link as

$$r = \left(\frac{P_t}{P_{min}}\right)^{1/\alpha} \quad (11)$$

Furthermore, the mobile users usually try to connect to the closest FAP by sending a call request. The request may be accepted or rejected according to the available capacity in the FAP or the users are not inside the FAP coverage's range [14]. The rejected requests may be served by the overlay macrocell. Without loss of generality, we are not focusing on capacity issue in this study. We consider an OAM as access method to reduce the cross-tier interference between femtocell and macrocell usage [15]. Thereby in our analysis, the mobile user is referred to femtocell user (FUE) or macrocell user (MUE). Therefore, the probability of a given user is fully connected ($P_C$) to the FAP is defined as the ratio:

$$P_C = \frac{D_{f,active}}{D_u} \qquad (12)$$

where $D_{f,active}$ denote to the density of active FAPs or defined as the density of active channels in a respective FAP. Therefore, $P_C$ can be approximated as follows [14]:

$$P_C = \frac{D_f}{D_u}\left[1 - \exp\left(-\frac{D_u}{D_f}\left(1 - \exp(-D_f \cdot \pi r^2)\right)\right)\right] \qquad (13)$$

Additionally, we study the effect of interference between the active FAP and mobile users in terms of outage probability [16], [17] and [18]; we consider the signal to interference (SIR) is given by:

$$SIR = \frac{P_t \times r_0^{-\alpha}}{N_0 + \sum_{i \in I} P_t \times r_i^{-\alpha}} \qquad (14)$$

where $r_0$ represents to the distance between the reference FAP and its respective mobile user, I is the set of the interferer users, $N_0$ represents the noise power and $r_i$ is the distance between the $i^{th}$ interferer at the reference FAP. Thereby, the outage probability ($P_{outage}$) is defined as the probability of FAP experiences a SIR less than or equal to a given threshold $\gamma$[19], i.e.

$$P_{outage} = \Pr\{SIR \leq \gamma\} \qquad (15)$$

Hence, $P_{outage}$ can be expressed as:

$$P_{outage} = 1 - \frac{D_f \cdot \left(1 - \exp-\left(D_{f,active}\gamma^{2/\alpha} + D_f\right) \cdot \pi r^2\right)}{\left(D_{f,active}\gamma^{2/\alpha} + D_f\right)\left[1 - \exp\left(-D_f \cdot \pi r^2\right)\right]} \qquad (16)$$

The evaluation of the minimum value on the outage probability is occurred at the SIR threshold $\gamma$. Under the usual assumption in interference-limited networks, the noise power is negligible and the overall interference power is treated as Gaussian noise, hence the minimum SIR ratio $\gamma$ required to guarantee a given spectral efficiency $\eta$ (bits/sec. Hz) at each active link can be calculated by using the Shannon's capacity formula, which yields

$$\gamma = 2^\eta - 1 \qquad (17)$$

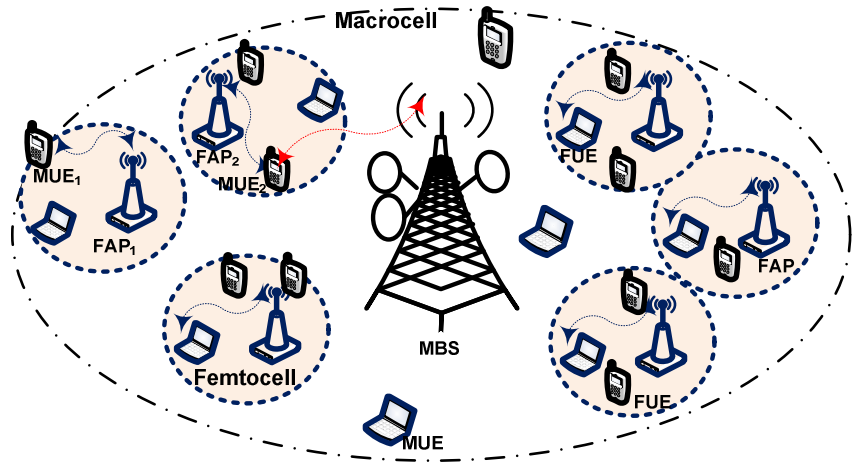

Figure 4. Femtocell/Macrocell mobile networks.

International Journal of Wireless & Mobile Networks (IJWMN) Vol. 4, No. 6, December 2012

## 6. NUMERICAL RESULTS

In this section, we present the numerical results of mobile user connectivity in femtocell/macrocell networks. The proposed analytical model has been the basis for the implementation of network model. Different input parameters for the network model are setting according to the real cellular networks. With the aid of Mathematica packages, we obtain some numerical solution and graphical illustrations for all metrics of interest, i.e. the probability of connectivity outage probability, and spectral efficiency.

### 6.1. Mobile user connectivity Probability

The user connectivity probability given by (9) is function of mobility factor ($\beta$), communication range ($r$) and number of femtocells ($N_f$). We evaluate the behavior of the proposed model under two different number of femtocells $N_f$ =100 and $N_f$ =10. Fig.5 and Fig.6 illustrate the probability of mobile user connectivity. As figures indicate, the connectivity probability increases by increasing the communication range especially when the mobility factor ($\beta$) is varied from 1 towards -1, i.e., the mobile users moved from outside to inside of femtocell. In contrast, the connectivity probability decreases when the mobility factor is varied from -1 towards 1. Also, we can observe that the probability of connectivity is increased when the number of femtocells is high. This is due to when the macrocell is highly dense with femtocells, the probability of mobile user connectivity increases. In sequel, Fig.7 concludes the above discussion.

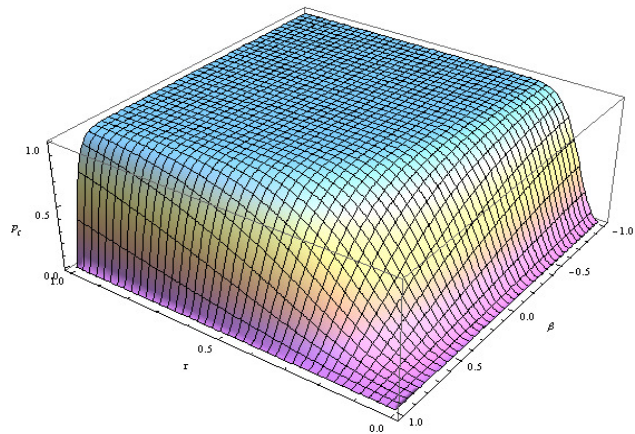

Figure 5. $P_c$ versus $\beta$ and $r$ at $N_f$ =100.

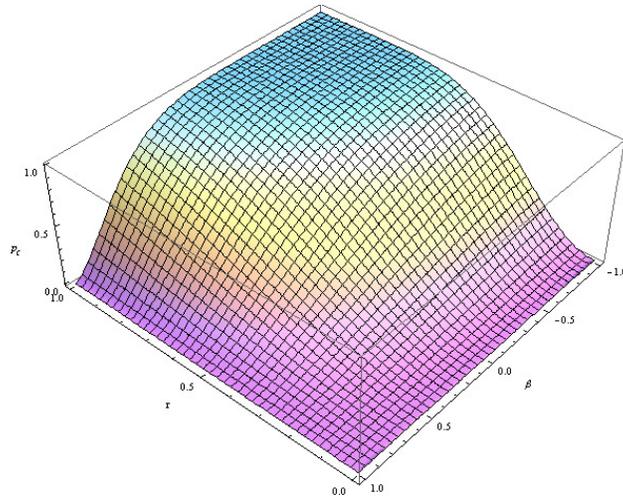

Figure 6 $P_c$ versus $\beta$ and $r$ at $N_f$ =10.

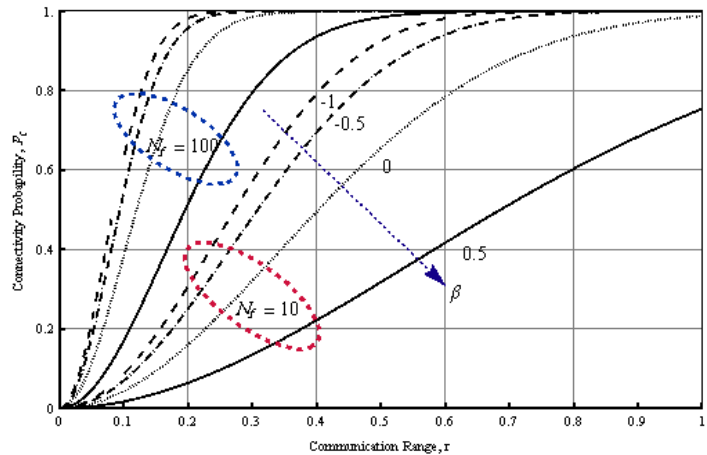

Figure 7 $P_c$ versus $r$ at different values of $\beta$ and $N_f$.

### 6.2. Outage probability and spectral efficiency

In this section, we study the performance of the connectivity probability ($P_C$), outage probability ($P_{outage}$) and the spectral efficiency ($\eta$) at different values of femtocell density ($D_f$) and user density ($D_u$).

We present the results of $P_C$ as a function of the density of users ($D_u$) at different values of femtocells densities ($D_f$) and at fixed communication range as shown in Fig.8. Femtocell density is produced different instances of the problem. As the figure indicates when the density of femtocells increases the user connectivity probability increases. However, when the density of users ($D_u$) increases, the user connectivity probability decreases. This is due to when the number of users which access the same FAP are increased, i.e., increasing the number of the rejected users, hence the user connectivity probability is decreased. In other word, the performance of femtocell/macrocell networks is significantly improved with increasing the density of femtocells.

Fig.9 shows the outage probability versus the density of femtocells at different user densities in femtocell/macrocell networks. We assumed the desired spectral efficiency in the network is $\eta$=2bits/s.Hz and the communication range is computed as in (10) at $\alpha$=4, $P_t$=1, $P_{min}$=10. Obviously, $P_{out}$ is steadily increases at lower density of femtocells, while $P_{out}$ decreases when the density of femtocells increased. Also, we observe $P_{out}$ is significantly decreased after $D_f$=2 FAPs/m² for different values of users density.

On the other hand, we study the performance of spectral efficiency ($\eta$) in femtocell/macrocell networks for achieving a given threshold value of outage probability ($P_{out}$). Fig. 10 illustrates the spectral efficiency ($\eta$) in bits/s.Hz when the user density increasing. As expected, the spectral efficiency decreases when the user density increases, this is due to increasing the number of users that used transmission channel. Obviously, at different threshold values of outage probability, we can get different plots of $\eta$. However, the performance of spectral efficiency is increased to achieve the desired $P_{out}$. In other word, the plots of $\eta$ are increased when the required $P_{out}$ is high. This is probably occurred when the number of outage users increased, the channel sends with high capacity (i.e. high $\eta$) and at the same time it satisfies the desired $P_{out}$.

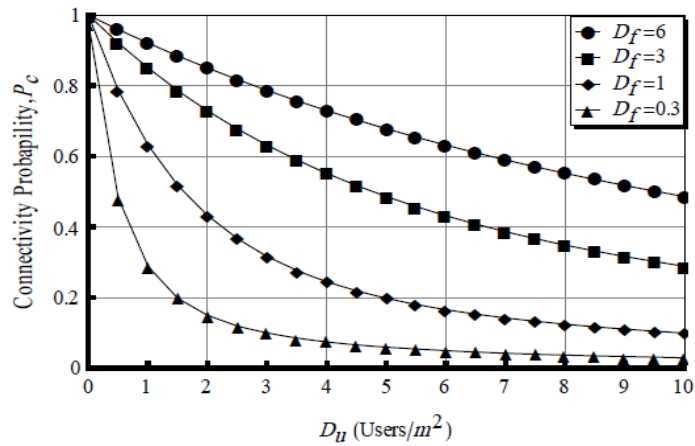

Figure 8. $P_c$ versus $D_u$ at different values of $D_f$.

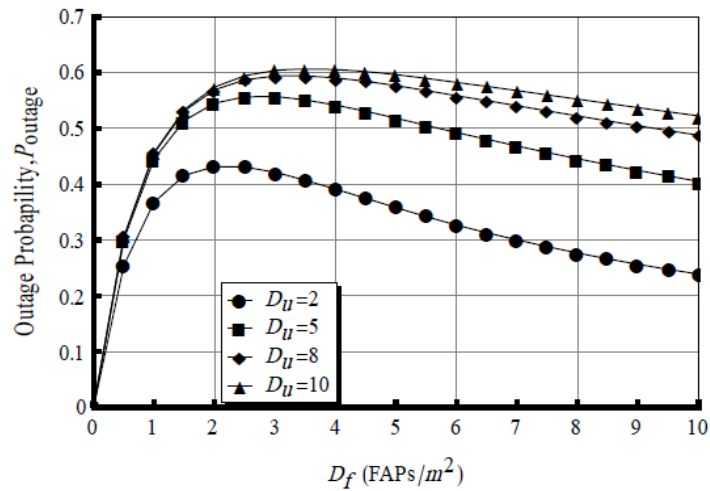

Figure 9. $P_{outage}$ versus $D_f$ at different values of $D_u$.

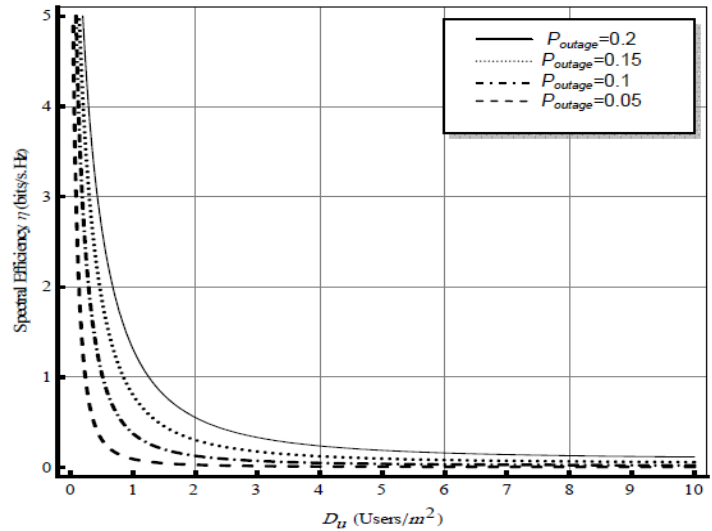

Figure 10.  $\eta$ versus $D_u$ at different threshold $P_{outage}$.

*To summarize, our analysis shows that, the probability of mobile user connectivity depends on different parameters such as communication range, femtocell density, and user density. It is shown that large number of femtocells in macrocell guaranteed feasible connectivity probability in Femto/Macro cellular networks.*

## 7. CONCLUSIONS

This paper presented an analytical model framework for computing the probability of mobile user connectivity in femtocell/macrocell networks. The performance of the connectivity probability is measured in terms of communication range, mobility factor, user density and femtocell density. In addition, we examined the performance of outage probability and spectral efficiency in such network.  Our studies demonstrated that the computing of user connectivity probability is essentially efficient during planning the Macrocellular networks integrated with Femtocellular networks.

## ACKNOWLEDGMENTS

This work was supported by the Deanship of Scientific Research in Salman Bin Abdulaziz University, Kingdom of Saudi Arabia under Grants No.20/أ/1432.

## REFERENCES

[1]  P. Lin, J. Zhang, Y. Chen, and Q. Zhang, "Macro-Femto Heterogeneous Network Deployment and Management: From Business Models to Techinical Solutions," IEEE Wireless Comm. Magazine, Vol. 18, No.3, June. 2011, pp 64 -70.

[2]  V. Chandrasekhar, J. Andrews, and A. Gatherer, "Femtocell networks: a survey," IEEE Comm. Magazine, Vol. 46, No.8, Sept. 2008, pp 59 - 67.

[3]  Guvenc I.,  M. Jeong, Watanabe F., and  Inamura H.," A hybrid frequency assignment for femtocells and coverage area analysis for co-channel operation" IEEE Comm. Letters, Vol. 12, No. 12, Dec. 2008, pp.880-882.

[4]  X. Chu, Y. Wu, L. Benmesbah, and W. Kuen Ling "Resource Allocation in Hybrid Macro/Femto Networks" in Proc. IEEE WCNC Workshop, April 2010, pp. 1-6.

[5]  Yongsheng S,   MacKenzie A., DaSilva L.A, Ghaboosi K., and  Latva-aho M.," On Resource Reuse for Cellular Networks with Femto- and Macrocell Coexistence" Proc. IEEE Globecom, Dec. 2010, pp.1-6.

[6] C. Han Ko and H. Yu Wei, "On-Demand Resource-Sharing Mechanism Design in Two-Tier OFDMA Femtocell Networks," IEEE Trans. on Vehicular. Tech., Vol. 60, No. 3, March 2011, p.p.1059-1071.
[7] Chandrasekhar V. and Andrews J, "Uplink capacity and interference avoidance for two-tier femtocell networks," IEEE Trans. Wireless Comm, Vol. 8 No. 7, July 2009, pp. 3498 – 3509.
[8] Ardian U., Robert B. and Melvi U., "Handover procedure and decision strategy in LTE-based femtocell network," Springer Telecommunication Systems, Online First, Sept. 2011.
[9] Jung-Min M. and Dong-Ho C., "Novel Handoff Decision Algorithm in Hierarchical Macro/Femto-Cell Networks" in Proc. IEEE WCNC, April 2010, pp. 1-6.
[10] A. Golaup, M. Mustapha, and L. Boonchin," Access control mechanisms for femtocells,"," IEEE Commun. Mag., Vol.48, No.1, January 2010, pp. 33-39.
[11] Feilu L., Bala E., Erkip E., and Rui Y.," A framework for femtocells to access both licensed and unlicensed bands" Proc. IEEE WiOpt, May 2011, pp.407-4011.
[12] Han-Shin J., Ping X. and Andrews J.," Downlink Femtocell Networks: Open or Closed?" Proc. IEEE ICC, June 2011, pp.1-65.
[13] M. Haenggi, J.G. Andrews, F. Baccelli, O. Dousse, and M. Franceschetti, "Stochastic geometry and random graphs for the analysis and design of wireless networks," IEEE Selected Area in Comm Journal, Vol. 27, No. 7, Sept. 2009, pp 1029 – 1046.
[14] Nardelli P.H.J., Cardieri P. and Latva-aho, "Efficiency of Wireless Networks under Different Hopping Strategies," IEEE Trans. Wireless Comm, Vol. 11, No. 1, January 2012, pp 15 –20.
[15] Tarasak, P., Quek T., and Chin F. "Uplink Timing Misalignment in Open and Closed Access OFDMA Femtocell Networks" IEEE Comm. Letters, Vol. 15, No. 9, Sept. 2011, pp. 926 – 928.
[16] V. Chandrasekhar, M. Kountouris, and J. G. Andrews, "Coverage in multi-antenna two-tier networks," IEEE Trans. Wireless Comm.,Vol. 8, No. 10, , Oct. 2009, pp. 5314–5327
[17] femtoforum.org/femto,"Interference Management in UMTS Femtocells," Dec. 2008.
[18] Ngo D., Le L., Le-Ngoc T., Hossain E., and Kim D., "Distributed Interference Management in Two-Tier CDMA Femtocell Networks," IEEE Trans. Wireless Comm., In press, No.99, 2012, pp. 1-11.
[19] K. Youngju, L. Sungeun, and H. Daesik, "Performance Analysis of Two-Tier Femtocell Networks with Outage Constraints," IEEE Trans. Wireless Comm.,Vol. 8, No. 9, Sept. 2010, pp. 5314–5327.
[20] R. Guillaume, A. Lada, L. David, C. Chia-Chin and J. Zhang," Self-Organization for LTE Enterprise Femtocells" Proc. IEEE Globecom Workshop on Femtocell Networks, Dec. 2010, pp.674-678.
[21] 3GPP Tech. Rep. 25.820, v. 8.2.0, Sept. 2008

## Authors Biographies

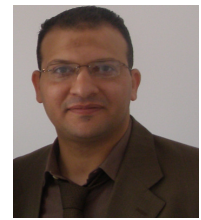

**Saied M. Abd El-atty** received the B.S. and M.S. degrees from Menoufia University, Faculty of Electronic Engineering, in 1995 and 2001, all in Electronics & Communications Engineering respectively. He received the PhD degree in Wireless Communication Networks from University of Aegean (UOA) at the Information and Communication Systems Engineering Department, Greece, Samos in 2008. He is a member of the faculty members in the department of Electronics and Electrical Communication at Faculty of Electronic Engineering, Menouf, Egypt. Currently, he is working as assistant professor in Salman Bin Abulaziz University, KSA. He is the head of computer science and information department in Science College. Dr. Saied's current research interests include design, analysis, and optimization of wireless mobile communication networks and vehicular networks as well as cognitive radio systems.

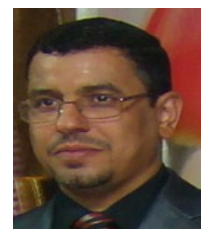

**Zakaria M. M. Gharsseldien** received his B.Sc. degree, M.S. degrees, and Ph.D. degree in Applied Mathematics from Al-Azhar University, Cairo, Egypt, in 1992.,1998, and 2003, respectively. He worked as a Lecturer in Department of Mathematics, Faculty of Science, Cairo, Al-Azhar University from 2003 to 2007. He is currently an Assistant Professor with the Department of Mathematics, Faculty of Arts and Science (Wadi Addwassir), Salman Bin AbdulAziz University from 2007. His current research interests is mathematical modeling in biology, medicine, and wireless communications, biomathematics, and bio fluid mechanics.